# Bulk Assembly of Organic Metal Halide Nanoribbons

Sujin Lee[†], Rijan Karkee[‡], Azza Ben-Akacha[†], Derek Luong[l], J.S. Raaj Vellore Winfred[†], Xinsong Lin[†], David A. Strubbe[‡,*], Biwu Ma[†,*]

[†] Department of Chemistry and Biochemistry, Florida State University, Tallahassee, Florida 32306, United States
[‡] Department of Physics, University of California, Merced, Merced, CA 95343, United States
[l] Department of Biology Science, Florida State University, Tallahassee, Florida 32306, United States

*Organic metal halide hybrids, nanoribbons, one dimensional structure, exciton self-trapping, dual emission*

**ABSTRACT:** Organic metal halide hybrids with low-dimensional structures at the molecular level have received great attention recently for their exceptional structural tunability and unique photophysical properties. Here we report for the first time the synthesis and characterization of a one-dimensional (1D) organic metal halide hybrid material, which contains metal halide nanoribbons with a width of three octahedral units. It is found that this material with a chemical formula $C_8H_{28}N_5Pb_3Cl_{11}$ shows a dual emission with a photoluminescence quantum efficiency (PLQE) of around 25% under ultraviolet (UV) light irradiation. Photophysical studies and density functional theory (DFT) calculations suggest the coexisting of delocalized free excitons and localized self-trapped excitons in metal halide nanoribbons leading to the dual emission. This work shows once again the exceptional tunability of organic metal halide hybrids that bridge between molecular systems with localized states and crystalline ones with electronic bands.

Metal halide perovskites and perovskite-related organic-inorganic hybrid materials have emerged as an important class of functional materials for a wide range of applications, including solar cells, light emitting diodes (LEDs), scintillators, etc.[1-6] The exceptional structural tunability of this class of hybrid materials via the control of both organic and metal halide components has led to the development of a variety of low dimensional structures at the molecular level, ranging from quasi-two dimensional (2D) to layered-2D,[7-16] corrugated-2D,[17-22] one-dimensional (1D),[23-34] and zero-dimensional (0D) structures.[35-36] Due to the isolation of metal halides by organic components, different degrees of electronic band formation and structural distortion can be achieved in these materials, which exhibit unique optical and electronic properties different from those of 3D metal halide perovskites.[36-37] For instance, narrowband emissions with small Stokes shifts from quantum-confined free excitons (FEs) are obtained in many quasi-2D and layered-2D metal halide perovskites[7-16] while strongly Stokes-shifted broadband emissions from localized self-trapped excitons (STEs) are recorded in most 1D and 0D organic metal halide hybrids.[33, 36] For a few corrugated-2D and 1D organic metal halide hybrids, the coexistence of FEs and STEs can produce multiband white emissions.[17-21, 24]

While early reports on 1D organic metal halide hybrids date back to the 1990s,[23] they have attracted great attention since 2017 with the discovery of 1D $C_4N_2H_{14}PbBr_4$, which exhibits bluish white emission with PLQEs of up to 20%.[24] Dozens of 1D organic metal halide hybrids have been reported since then with improved PLQEs.[25-30] For instance, Gautier et al. reported white emitting 1D (TDMP)PbBr$_4$ with a PLQE of around 45%.[26] More recently, a PLQE of around 60% was recorded in yellowish white-emitting 1D $[C_4N_2H_{12}]_3[PbBr_5]_2 \cdot 4DMSO$.[30] The rich chemistry of organic metal halide hybrids has enabled synthesis of diverse 1D structures containing corner-, edge-, and face-sharing metal halides. In addition to 1D linear chains, corrugated and tubular 1D structures have been assembled using metal halide building blocks.[31-32] The different 1D structures lead to distinct optical properties, with some showing dual emissions from both FEs and STEs and others exhibiting strongly Stokes-shifted broadband emissions from STEs only. The recent advances in 1D organic metal halide hybrids suggest that there is ample room to develop new types of 1D structures.[33] One classic 1D structure that has not been well explored for metal halides is 1D ribbons. Like using benzene rings as building block for the formation of 1D graphene nanoribbons,[38-42] we hypothesize that 1D metal halide nanoribbons could be prepared using metal halide polyhedra as building block.

Here we report for the first time the synthesis and characterization of a 1D organic metal halide hybrid, $C_8H_{28}N_5Pb_3Cl_{11}$, containing metal halide nanoribbons with a width of three octahedral units. In this unique 1D hybrid, the corner- and edge-sharing octahedral lead chloride chains $(Pb_3Cl_{11}{}^{5-})_\infty$ are encompassed by long multiply charged organic tetraethylenepentammonium cations (TEPA$^{5+}$, $C_8H_{28}N_5{}^{5+}$). Greenish-white light emission with peaks at around 420 nm and 540 nm is observed under UV light irradiation (365 nm) at room temperature, with a PLQE of around 25%.

$C_8H_{28}N_5Pb_3Cl_{11}$ single crystals were prepared by gradually cooling a hydrochloric acid aqueous solution of tetraethylenepentamine pentahydrochloride (TEPA·5HCl) and lead chloride (PbCl$_2$) (Figure 1a). The crystal structure of colorless thin plate-like single crystals was determined by single-crystal X-ray diffraction (SCXRD), which shows a structure with orthorhombic space group Pbca. Detailed structural analysis can be found in the Supporting Information (Table S1). The uniformity and structural phase consistency of 1D $C_8H_{28}N_5Pb_3Cl_{11}$ single crystals were confirmed by powder XRD (Figure S1). Although $C_8H_{28}N_5Pb_3Cl_{11}$ single crystals were found to contain trace amounts of water molecules, as confirmed by thermogravimetric analysis (Figure S2) and elemental analysis (see details in the Supporting information), computational studies suggest that such O-containing impurities have little-to-no effect on their structural properties, and a minor effect on the optical properties, primarily an increase in absorption at shorter wavelengths for x-polarization (Figure S3).

The novel 1D structure at the molecular level is depicted in Figure 1b, in which anionic metal halide nanoribbons $(Pb_3Cl_{11}^{5-})_\infty$ are isolated and surrounded by multi-charged organic cations $C_8H_{28}N_5^{5+}$. For an individual metal halide nanoribbon (Figure 1c), metal halide octahedra $(PbCl_6^{4-})$ are connected via both corner-sharing in the long $a$-direction and edge-sharing in the short $c$-direction, unlike corner-sharing only as in typical layered-2D perovskites.[14-16] The width of an individual nanoribbon with three metal halide octahedral units is about 0.92 nm, which is significantly smaller than the typical Bohr exciton radius of metal halide perovskites (~7 nm),[43] suggesting strong quantum confinement in two dimensions. As shown in many previous studies, the geometry and charge of organic cations play an essential role in controlling the formation of low dimensional metal halides.[44-47] The nanoribbon structure here is indeed enabled by organic cations $C_8H_{28}N_5^{5+}$, where the length of an organic cation (~14 Å) is similar to that of three metal halide octahedral units (~12 Å). Also, the distance between nearby ammonium groups (~3.8 Å) is similar to the length between two chloride atoms along the octahedral edge (~3.8-4.2 Å), allowing ammonium groups to bind closely with chloride atoms to form a layer of organic cations on metal halide nanoribbons.

Structural distortion is an important characteristics of metal halides, which affects the formation of STEs and subsequently the photophysical properties. For 1D $C_8H_{28}N_5Pb_3Cl_{11}$, three types of metal halide octahedra $(PbCl_6^{4-})$ with different degrees of structural distortions are present in an individual nanoribbon (Figure 1d). To quantify the structural distortions, we have calculated the distortion of each $PbXCl_6^{4-}$ (X = 1, 2, and 3) octahedron according to the formula $\Delta_{oct} = \frac{1}{6}\sum_i \left(\frac{d_i - d_{avg}}{d_{avg}}\right)^2$,[48] in which $d_i$ is the distance of each Pb-Cl bond and $d_{avg}$ is the average Pb-Cl bond distance. $Pb1Cl_6^{4-}$ is located in the middle of the ribbon, linking both $Pb2Cl_6^{4-}$ and $Pb3Cl_6^{4-}$. The Pb1-Cl bond lengths range from 2.185 to 2.957 Å and $\Delta_{oct}$ of $Pb1Cl_6^{4-}$ is 3.0 × $10^{-4}$. $Pb2Cl_6^{4-}$ and $Pb3Cl_6^{4-}$, located on both sides of the metal halide nanoribbon along the $a$-axis, are highly distorted compared to $Pb1Cl_6^{4-}$, giving $\Delta_{oct}$ of 21.0 × $10^{-4}$ and 20.0 × $10^{-4}$, respectively, with Pb-Cl bond lengths ranging from 2.695 to 3.112 Å in $Pb2Cl_6^{4-}$ and from 2.728 to 3.085 Å in $Pb3Cl_6^{4-}$. Metal halide distortions of this magnitude have been found to lead to the formation of STEs.[22, 32, 34, 37]

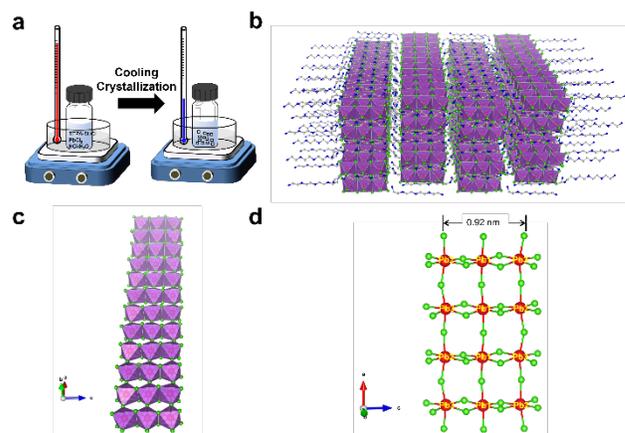

Figure 1. (a) Synthetic scheme for the material preparation; (b) crystal structure of $C_8H_{28}N_5Pb_3Cl_{11}$ (Grey spheres, carbon atoms; blue spheres, nitrogen atoms; red spheres, lead atoms; green spheres, chloride atoms; hydrogen atoms and oxygen atoms were hidden for clarity); (c) view of an individual lead chloride nanoribbons; (d) ball-and-stick model of a lead chloride nanoribbon.

The photophysical properties of $C_8H_{28}N_5Pb_3Cl_{11}$ are characterized via both steady-state and time-resolved spectroscopies. As shown in Figure 2a, 1D $C_8H_{28}N_5Pb_3Cl_{11}$ single crystals are colorless under ambient light and exhibit greenish-white emission upon ultraviolet (UV) irradiation (365 nm). The excitation and emission spectra of $C_8H_{28}N_5Pb_3Cl_{11}$ are shown in Figure 2b. A dual emission with a PLQE of 25% (Figure S4) is recorded, whereas the high-energy emission band peaked at around 420 nm has a full width at half maximum (FWHM) of ~60 nm (~ 0.46 eV) and the low-energy emission band peaked at around 540 nm has a FWHM of ~100 nm (~ 0.41 eV). The excitation spectra (Figure 2b) for peak emissions at 420 nm and 540 nm are found to be almost identical, and located at the absorption band edge (Figure S5), suggesting that both emissions originate from the same initially excited state. Room-temperature time-resolved photoluminescence spectra for peak emissions at 420 nm and 540 nm are shown in Figure 2c-d, giving decay lifetimes of ~1.8 ns and ~250 µs, respectively. Considering that similar dual emissions have previously been observed in corrugated-2D and 1D organic metal halides where FEs and STEs coexist,[22, 24] it is reasonable to attribute the high-energy emission to FEs and the low-energy emission to the STEs in this 1D $C_8H_{28}N_5Pb_3Cl_{11}$. Unlike many previously reported systems, showing similar decay lifetimes for the emissions from FEs and STEs due to thermal equilibrium at RT, 1D $C_8H_{28}N_5Pb_3Cl_{11}$ shows completely different decay behaviors for FEs and STEs (Figure S6), suggesting that emitting states are not in thermal equilibrium in this system. In other words, at short times after the initial excitation, excitons are distributed between FEs and STEs without detrapping from STEs to FEs. This behavior is indeed similar to that observed in many molecular systems that show both fluorescence and phosphorescence.[49-51]

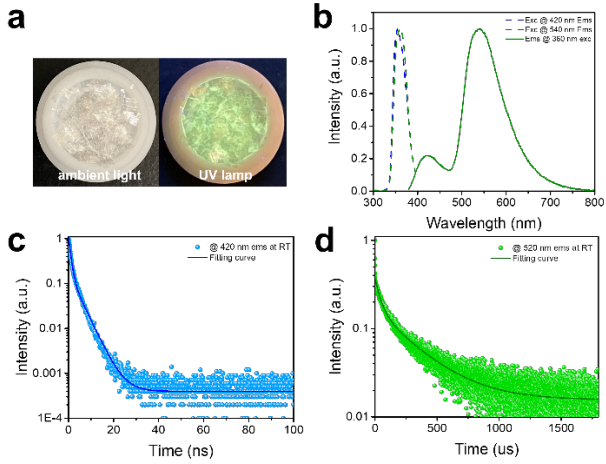

Figure 2. (a) Images of $C_8H_{28}N_5Pb_3Cl_{11}$ single crystals under ambient light (left) and under UV (right); (b) excitation and emission spectra of $C_8H_{28}N_5Pb_3Cl_{11}$ single crystals at room temperature; (c) photoluminescence decay for the emission peaked at 420 nm; (d) photoluminescence decay for the emission peaked at 540 nm.

To further clarify the intrinsic nature of the two emission bands, we have measured the dependence of emission intensity on excitation power density at room temperature (Figure 3a). The intensities of both emission bands show a linear dependence on the excitation power density, suggesting that both emissions are intrinsic properties rather than from defects.[52] Furthermore, to exclude an origin of emissions from permanent defects on the surface that often occur in conventional semiconducting materials such as CdSe,[53-54] we have compared the emission spectra of bulk single crystals and ground powder of $C_8H_{28}N_5Pb_3Cl_{11}$. The almost identical emissions from single crystals and ground powder (Figure S7) confirm no significant emission from surface defects. Low-temperature photophysical properties were also characterized for 1D $C_8H_{28}N_5Pb_3Cl_{11}$ at 77K. Like most low-dimensional organic metal halide hybrids,[55-56] only strongly Stokes-shifted broadband emission peaked at ~580 nm from STEs is recorded at 77K (Figure 3b) with a decay lifetime of around 4.6 μs (Figure 3c). Considering the distinct photophysical properties at RT and 77K, we believe the key photophysical processes are as depicted in Figure 3d. At room temperature, upon photoexcitation, 1D $C_8H_{28}N_5Pb_3Cl_{11}$ generates both FEs and STEs without detrapping, resulting in two emissions with different decay dynamics. At 77K, fast exciton self-trapping leads to a broadband emission from STEs only.

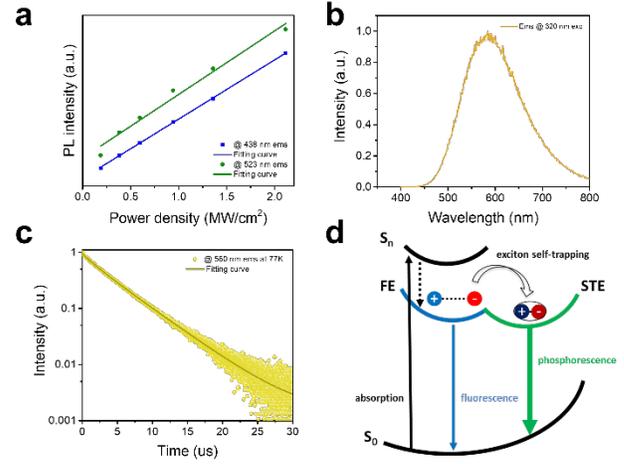

Figure 3. (a) Excitation power-dependent photoluminescence; (b) emission spectra of $C_8H_{28}N_5Pb_3Cl_{11}$ single crystals at 77K; (c) photoluminescence decay at 77K; (d) schematic photophysical processes in $C_8H_{28}N_5Pb_3Cl_{11}$.

To gain a better understanding of the structure-property relationship for 1D $C_8H_{28}N_5Pb_3Cl_{11}$, we have performed plane-wave DFT calculations using Quantum ESPRESSO[57] and the Perdew-Burke-Ernzerhof (PBE)[58] functional. Relaxation starting from the SCXRD structure ($a$ = 11.35 Å, $b$ = 15.85 Å, $c$ = 32.12 Å, $\alpha = \beta = \gamma = 90°$) gave lattice parameters in close agreement ($a$ = 11.32 Å, $b$ = 15.85 Å, $c$ = 32.15 Å, $\alpha = \beta = \gamma = 90°$), as in work on other hybrid perovskites.[59] The calculated electronic band structure of $C_8H_{28}N_5Pb_3Cl_{11}$ is shown in Figure 4a; the gap is direct and computed to be 3.24 eV. This is close to, though slightly less than, the experimental result from Tauc fitting (Figure S8), in accordance with the common cancellation of errors between neglect of spin-orbit coupling and quasiparticle corrections in hybrid perovskites.[60] The valence band maximum (VBM) has contribution mostly from $p$-orbitals of Cl and C, whereas at the conduction band minimum (CBM), $p$-orbitals of Pb dominate (Figure 4b, 4c). The electronic bands are dispersive along the Pb-Cl chain direction ($x$), with effective masses at the Γ point of $m^*_{VBM}$ = -0.34 $m_0$ and $m^*_{CBM}$ = 0.09 $m_0$, but are nearly flat along other two perpendicular directions ($y$: $m^*_{VBM}$ = -2.25 $m_0$, $m^*_{CBM}$ = 4.23 $m_0$; $z$: $m^*_{VBM}$ = -18.5 $m_0$, $m^*_{CBM}$ = 12.5 $m_0$) as shown in Figure 4a.

Additionally, we calculated the absorption through the random phase approximation (RPA) in BerkeleyGW code,[61] shown in Figure 4d. The first absorption peak is strong along the Pb-Cl chain ($x$-polarized), suggesting that the electronic transition in the band edges is coming from the Pb-Cl chain. For $x$-polarization, the strong transition is due to VBM-8 and VBM-7 to CBM+4 bands (very close in energy to VBM to CBM transitions). This calculated direct absorption peak matches well with our experimental absorption results (Figure S5, S8). Similarly for $y$-polarization, a transition from VBM-17 to CBM+47 bands dominates; and for $z$-polarization, from VBM-21 to CBM+27 bands. Furthermore, Figure 4d shows strong anisotropy in absorption. These calculations indicate that the FE emission comes from the recombination of excitons on the Pb-Cl chain. The experimental emission energy for both emitting states should be lower than the bandgap due to the strong exciton binding as usual in low-dimensional structures,[62] as well as Stokes shifts. We also constructed an analogous bulk structure containing 2D sheets of Pb-Cl atoms in plane (Figure S9), separated by organic molecules out of plane, and obtained a bandgap of 2.84 eV. The orbital nature of the band edges is similar to 1D $C_8H_{28}N_5Pb_3Cl_{11}$. This suggests a quantum confinement effect of 0.40 eV in this material as one goes from 2D to 1D.

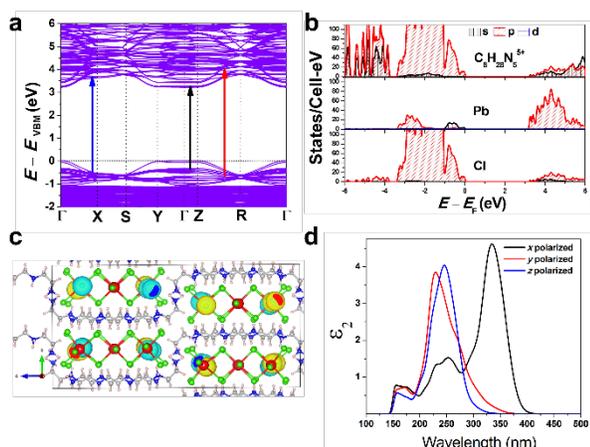

Figure 4. Density-functional theory simulations: (a) Electronic band structure, with arrows showing transitions polarized along $x$ (black), $y$ (red), and $z$ (blue) directions, where $x$ is the direction along the Pb-Cl chains; (b) partial density of states; (c) wavefunction of the conduction band minimum at Γ, localized along the Pb-Cl chains; (d) polarized absorption spectra.

In conclusion, we have developed a novel 1D organic metal halide hybrid ($C_8H_{28}N_5Pb_3Cl_{11}$) consisting of metal halide nanoribbons with a width of three octahedral units via cooling crystallization. 1D $C_8H_{28}N_5Pb_3Cl_{11}$ is found to exhibit a dual emission with a high energy emission peaked at 420 nm and a low energy broad emission peaked at 540 nm, due to the coexistence of free excitons and self-trapped excitons. Our work further advances research on low dimensional organic metal halide hybrids with a unique nanoribbon structure, which bridges between linear chain 1D structures and layered-2D structures. It shows once again the exceptional tunability of organic metal halide hybrids, which could lead to the further development of functional hybrid materials for many applications.

## ASSOCIATED CONTENT

**Supporting Information**. Syntheses and characterizations of $C_8H_{28}N_5Pb_3Cl_{11}$, and computational details and further results (PDF). DFT relaxed structures in XCrySDen format (ZIP). This material is available free of charge via the Internet at http://pubs.acs.org.

## AUTHOR INFORMATION

**Corresponding Author**
* Biwu Ma, bma@fsu.edu; David A. Strubbe, dstrubbe@ucmerced.edu

**Notes**
The authors declare no competing financial interests.

## ACKNOWLEDGMENT

The work is supported by the National Science Foundation (NSF) (DMR- 2204466). This work made use of the Rigaku Synergy-S single-crystal X-ray diffractometer which was acquired through the NSF MRI program (award CHE-1828362). Nanosecond-transient absorption measurement was performed on spectrometers supported by the National Science Foundation under Grant No. CHE-1531629. A portion of this research used resources provided by the Materials Characterization Laboratory (FSU075000MAC) at the FSU Department of Chemistry and Biochemistry. Computational work was supported by the Air Force Office of Scientific Research under award number FA9550-19-1-0236. Computational resources were provided by the Multi-Environment Computer for Exploration and Discovery (MERCED) cluster at UC Merced, funded by National Science Foundation Grant No. ACI-1429783.

Insert Table of Contents artwork here

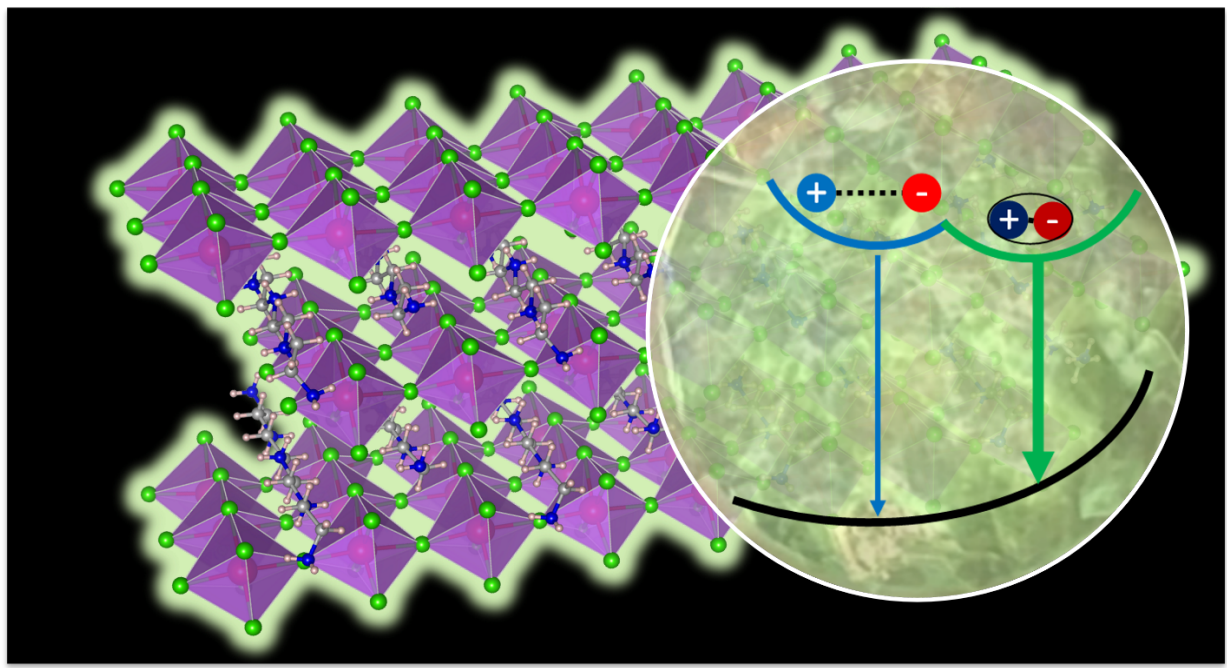

# Supporting Information

## Bulk Assembly of Organic Metal Halide Nanoribbons

Sujin Lee[†], Rijan Karkee[‡], Azza Ben-Akacha[†], Derek Luong[l], J.S. Raaj Vellore Winfred[†], Xinsong Lin[†], David A. Strubbe[‡,*], Biwu Ma[†,*]

[†] Department of Chemistry and Biochemistry, Florida State University, Tallahassee, Florida 32306, United States
[‡] Department of Physics, University of California, Merced, Merced, CA 95343, United States
[l] Department of Biology Science, Florida State University, Tallahassee, Florida 32306, United States

**Materials and Characterization**

**Materials.** Lead chloride ($PbCl_2$, 99.99 %), tetraethylenepentamine pentahydrochloride (98%), HCl (37%) were purchased from Sigma Aldrich and were used without further purification.

**Synthesis of $C_8H_{28}N_5Pb_3Cl_{11}$.** Lead chloride (0.40 mmol) and tetraethylenepentamine pentahydrochloride (0.13 mmol) were added to a vial with 5 mL of HCl and 5 mL of water. The mixture was stirred and heated at 100 °C for 1 hour. Then the solution was left to cool slowly to room temperature. Colorless thin platelike crystals were obtained. The crystals were washed with water and acetone and dried under reduced pressure. The yield was calculated at ~ 67 %. $C_8H_{28}N_5Pb_3Cl_{11}$: Anal, Calc. C, 7.97; H, 2.34; N, 5.81; Cl, 32.34 Found: C, 7.96; H, 2.33; N, 5.61; Cl, 32.07.

**Single crystal X-ray crystallography:** The single crystal structure was solved using Rigaku XtaLAB Synergy-S diffractometer equipped with a HyPix-6000HE Hybrid Photon Counting (HPC) detector and Cu microfocus sealed X-ray sources. The crystal was mounted in cryoloop with Paratone-N oil. The data were collected at 150.00(10) K. The structures were solved using

Olex2 software where XT[1] structure solution program using Intrinsic Phasing and the XL[2] refinement package using Least Squares minimization were employed in solving and refining the structures, respectively. Table S1 summarizes the refinement details and the resulting factors. A CIF has been deposited with the CCDC (2181734). VESTA was used as the crystal structure visualization software for the images presented in the manuscript.

**Powder X-ray Diffraction:** The PXRD analysis was performed on Rigaku MiniFlex powder X-Ray diffractometer with a copper anode-type generator (Cu-Ka, λ =0.154 nm, 40 kV, 15 mA) and D/teX Ultra detector. The diffraction pattern was scanned over the angular range of 5-50º (2θ) with a step size of 0.05 at room temperature. Simulated powder patterns were calculated in the Mercury software from the corresponding crystallographic information file from SCXRD experiment.

**Thermogravimetry Analysis:** TGA was conducted on a TA Instruments TGA 550 system. The sample was heated from room temperature to 250 ºC at a rate of 5 ºC·min$^{-1}$, then heated at a rate of 10 ºC·min$^{-1}$ to 800 ºC under an argon flux of 40 mL·min$^{-1}$.

**Photoluminescence Steady-State Studies:** Steady-state PL spectra of $C_8H_{28}N_5Pb_3Cl_{11}$ were measured at room temperature and at 77 K (liquid nitrogen was used to cool the sample) on a FS5 spectrofluorometer (Edinburgh Instruments).

**Absorption Spectrum Measurements:** Absorption spectra of $C_8H_{28}N_5Pb_3Cl_{11}$ were obtained at room temperature through a synchronous scan in an integrating sphere incorporated into the spectrofluorometer (FS5, Edinburgh Instruments).

**Time-resolved photoluminescence:** The lifetimes in the range of nanosecond were measured at room temperature using time-correlated single photon counting on a FS5 spectrofluorometer (Edinburgh Instruments) with data collection for 10000 counts. The samples were excited by an Edinburgh EPL-365 picosecond pulsed diode laser and the emission counts was monitored at corresponding emission maximums. The long lifetime in the range of microsecond was measured using the Edinburgh Instruments LP980-KS laser flash photolysis spectrometer using an Nd:YAG pump laser as the excitation source. The emission counts were monitored at the corresponding emission maximum, and the data was obtained to 10000 counts.

**Photoluminescence Intensity Dependence on Excitation Power Density:** The power-dependent PL intensity measurement was performed on an Edinburgh Instruments LP980-KS transient absorption spectrometer using a Continuum Nd:YAG laser (Surelite EX) pumping a Continuum optical parametric oscillator (Horizon II OPO) to provide 360 nm 5 ns pulses at 1 Hz. The pump beam profile was carefully defined using collimated laser pulses passed through an iris set to an area of 0.38 cm$^2$. The pulse intensity was monitored by a power meter (Ophir PE10BF-C), detecting the reflection from a beam splitter. Detection consisted of an Andor Intensified CCD (1024 × 256 element) camera, collecting a spectrum from 287 to 868 nm and gated to optimize PL collection (typically a 30−50 ns gate depending on the PL lifetime, starting immediately following the 5 ns laser pulse). Twenty collections were averaged at each power level with every laser pulse monitored to determine the average intensity. The PL was determined at the maximum of the PL emission curve.

**Photoluminescence Quantum Efficiencies:** The PLQE was obtained using a Hamamatsu Quantaurus-QY Spectrometer (model C11347-11) equipped with a xenon lamp, integrated sphere sample chamber, and CCD detector. The PLQE was calculated by using the equation: $\eta_{QE} = I_S/(E_R - E_S)$, in which $I_S$ represents the luminescence emission spectrum of the sample, $E_R$ is the spectrum of the excitation light from the empty integrated sphere (without the sample), and $E_S$ is the excitation spectrum for exciting the sample.

**Computational Methods:** Calculations used Quantum ESPRESSO version 6.4.1.[3] ONCV

pseudopotentials[5] from PseudoDojo[6] were used. A wavefunction energy cutoff of 816 eV was used, and a 3×2×1 half-shifted *k*-grid for self-consistent field (SCF) calculations. Forces and stresses were relaxed below $10^{-4}$ Ry/bohr and 0.1 kbar, respectively. Density of states calculations used a 9×6×3 half-shifted *k*-grid and a broadening of 0.05 eV. Optical absorption spectra computed with the random phase approximation (RPA) in BerkeleyGW used 100 occupied states and 150 unoccupied states, 10×2×2 k-point sampling, and 0.1 eV Gaussian broadening.

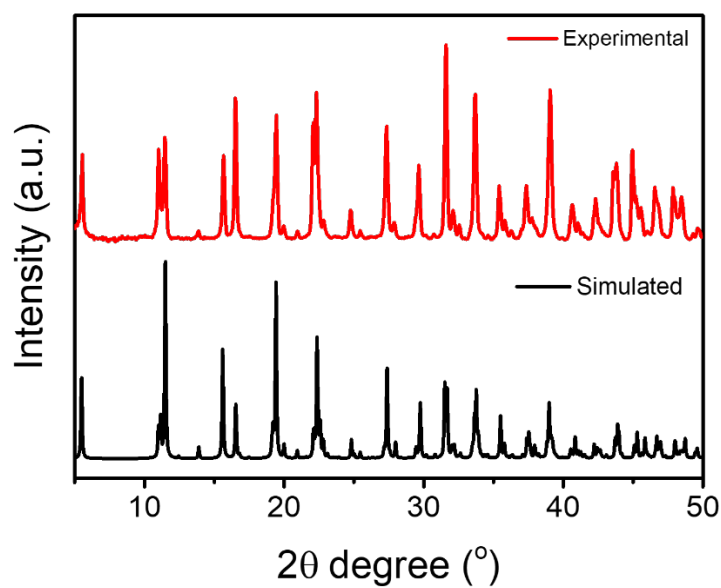

**Figure S1.** PXRD patterns of $C_8H_{28}N_5Pb_3Cl_{11}$ and its SCXRD simulated result.

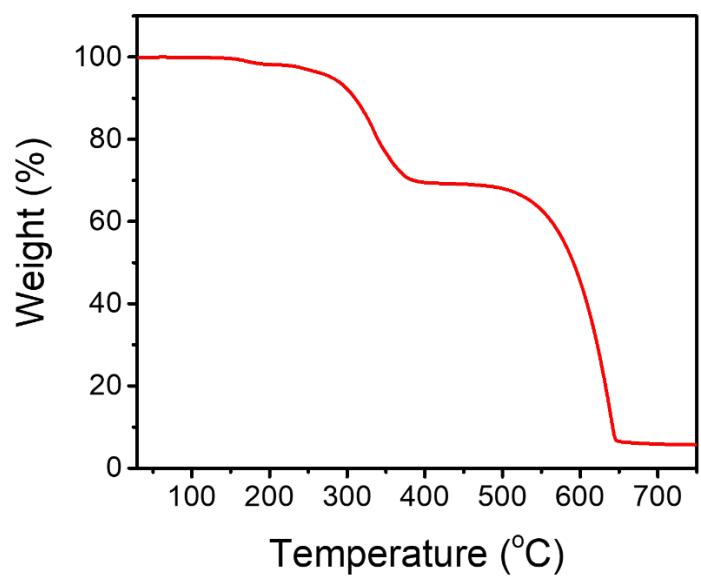

**Figure S2.** TGA analysis of $C_8H_{28}N_5Pb_3Cl_{11}$.

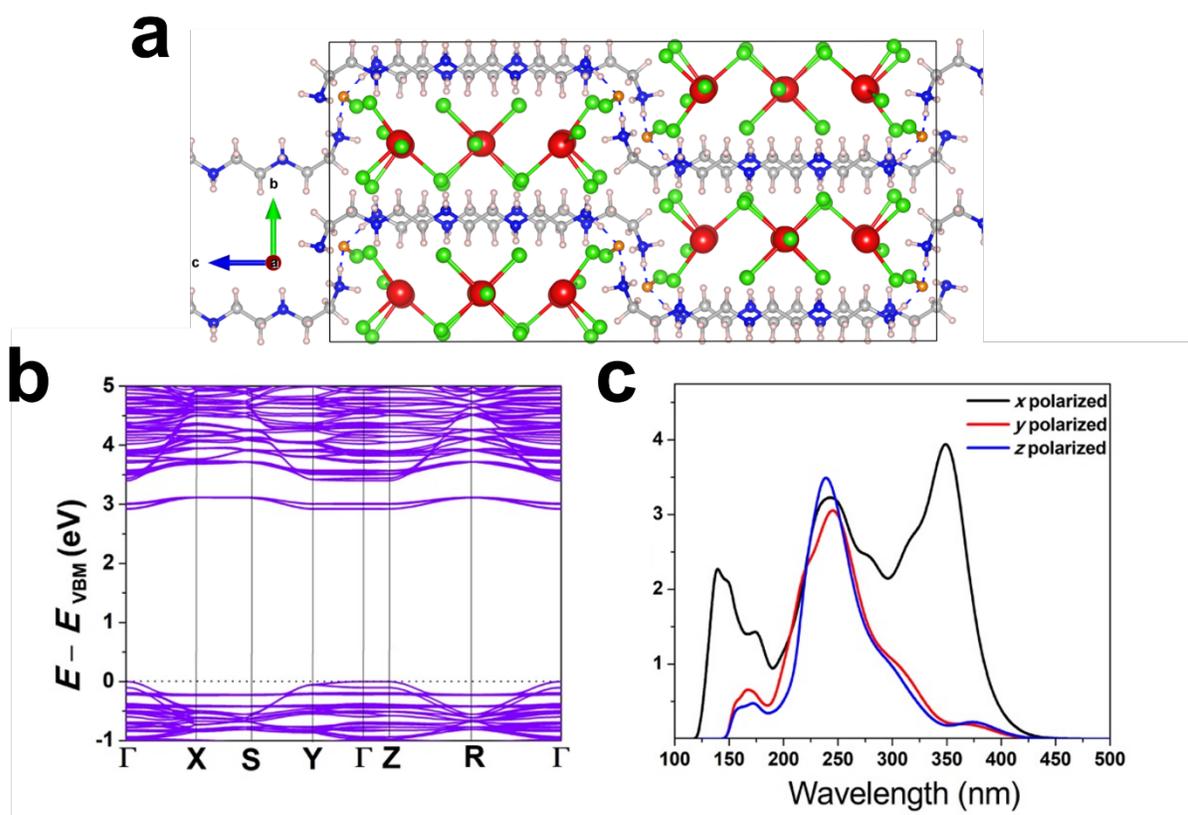

**Figure S3.** Calculations of structure from SCXRD containing an O atom, with formula unit

$C_8H_{28}Cl_{11}N_5OPb_3$. (a) Relaxed structure with lattice parameters $a = 11.32$ Å, $b = 15.92$ Å, $c = 32.15$ Å, $\alpha = \beta = \gamma = 90°$; minimally different from structure without O atom. The O atom (orange) makes a bond with Cl. (b) Bandstructure, showing a new flat conduction band in the gap, localized on the O-Cl bond. The next lowest conduction bands are on the Pb-Cl chain as in the structure without O, and they are pushed up slightly higher in energy. (c) Optical absorption spectrum, showing similar features to that without O, but with somewhat modified intensities of higher peaks. The O-Cl states, due to their localization, make little contribution.

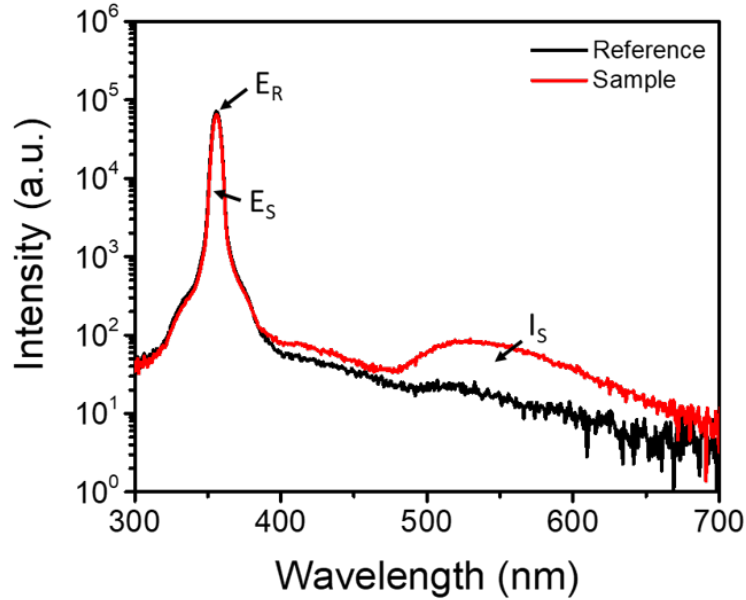

**Figure S4.** Excitation line of reference and emission spectrum of 1D lead chloride nanoribbons collected by an integrating sphere. The PLQE was calculated by the equation: $\eta_{QE} = I_S/(E_R - E_S)$.

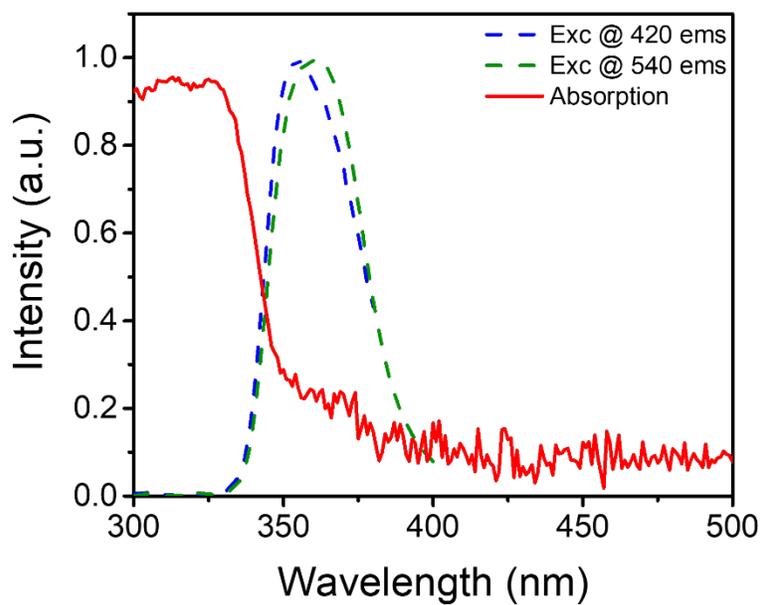

**Figure S5.** Absorption and excitation spectra at room temperature.

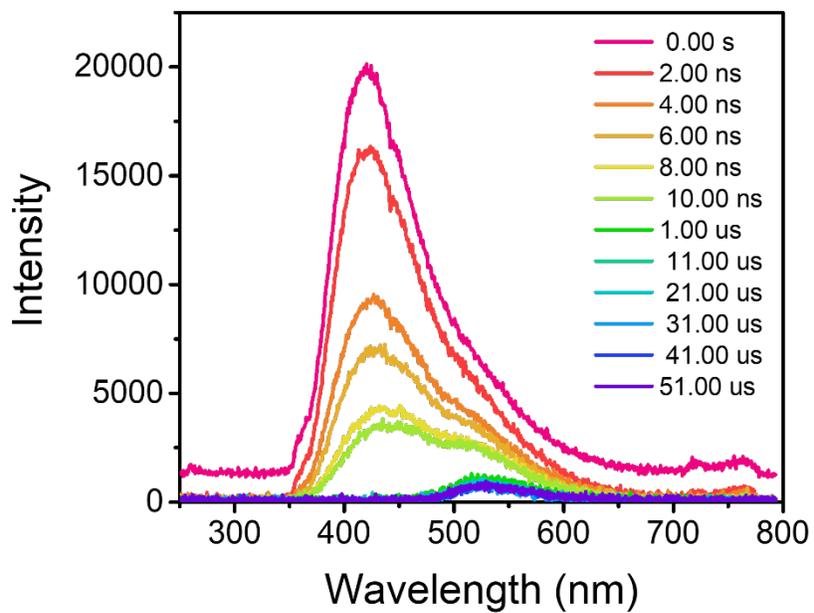

**Figure S6.** Emission spectra at a specific gate delay time.

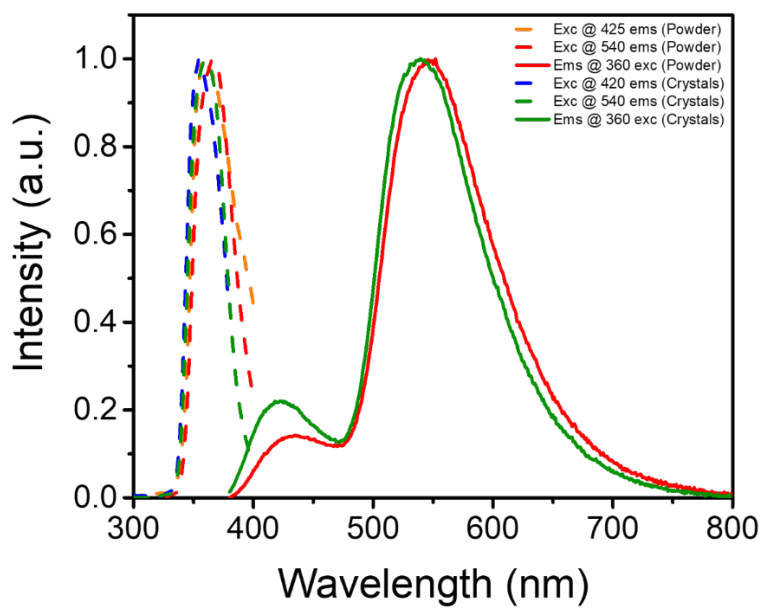

**Figure S7.** PL spectra of ground powder and single crystals at RT.

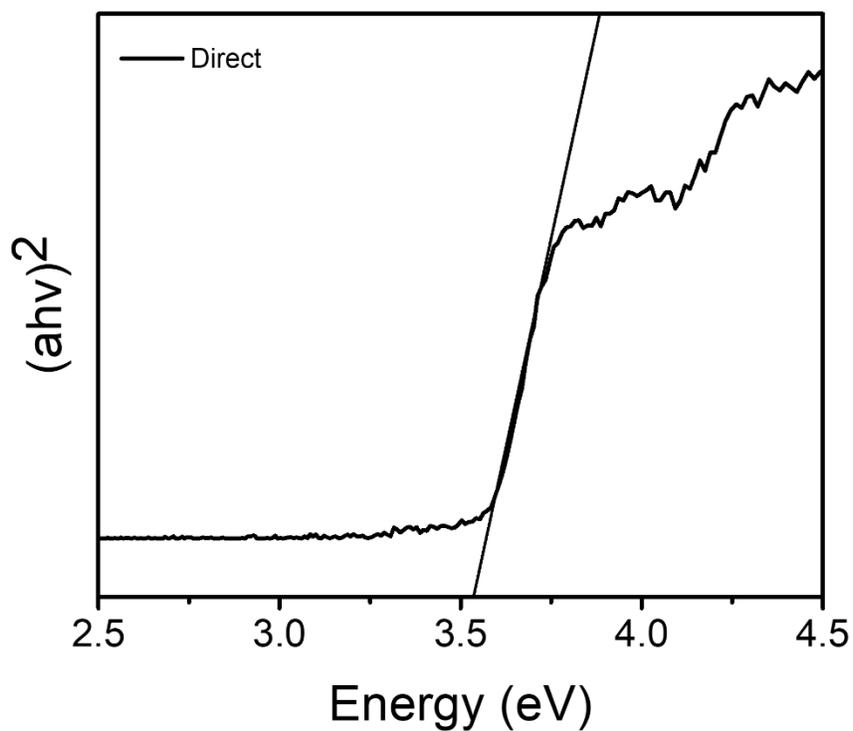

**Figure S8.** Tauc plot for the absorption spectrum of $C_8H_{28}N_5Pb_3Cl_{11}$ crystals.

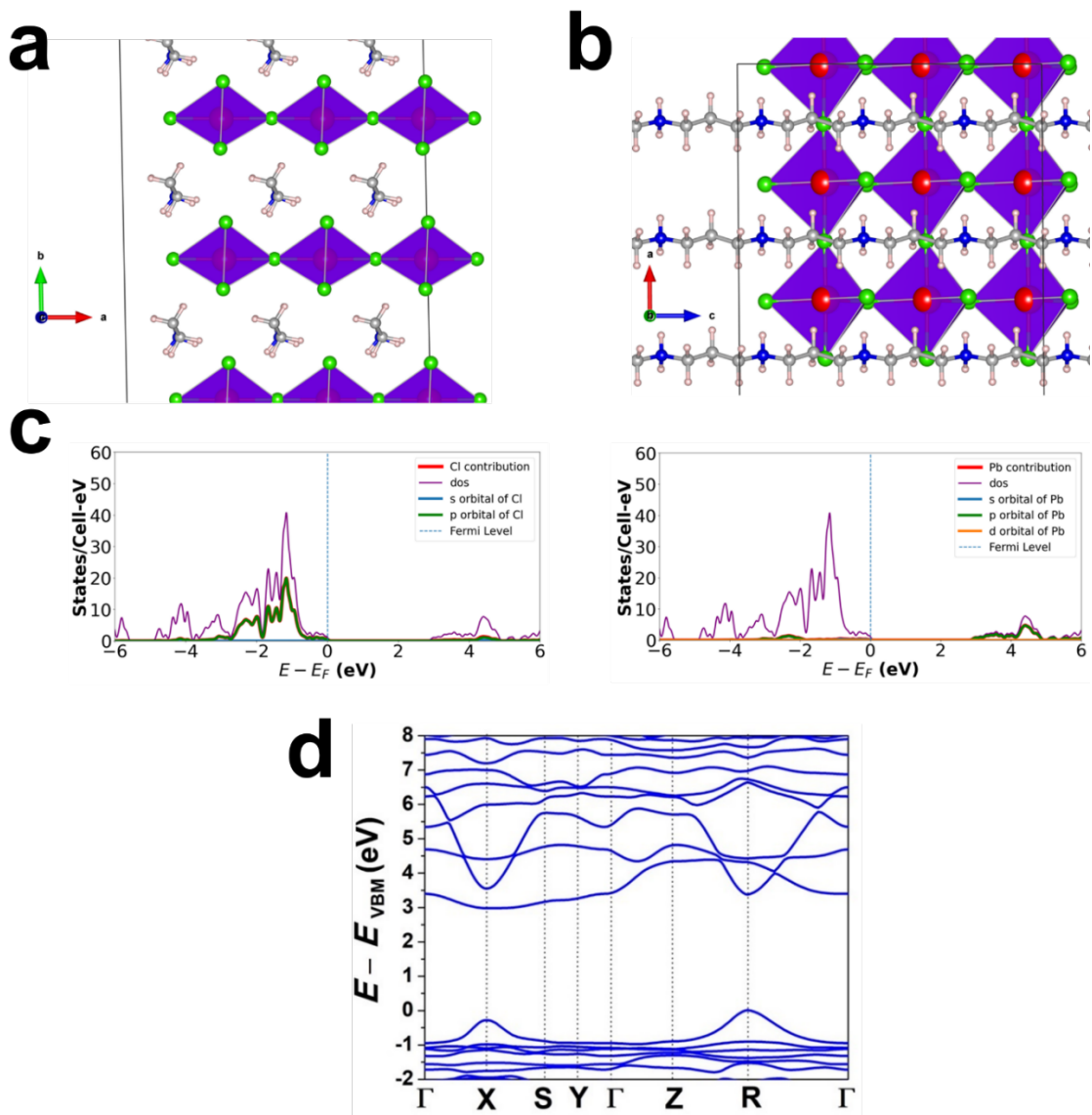

**Figure S9.** Calculations of hypothetical 2D sheet structure, corresponding to the 1D C$_8$H$_{28}$N$_5$Pb$_3$Cl$_{11}$, to show quantum confinement effects. The relaxed structure has lattice parameters $a$ = 5.67Å, $b$ = 7.92 Å, $c$ = 4.92 Å, $\alpha$ = 93.35°, $\beta$ = 90.33°, $\gamma$ = 91.15°. The unit cell contains 16 atoms with chemical formula C$_3$H$_8$NPbCl$_3$. The Pb-Cl sheets are in the *ac* plane, separated by polymeric cations. SCF calculations used a 4×2×4 half-shifted k-grid. (a) View showing separation between layers. (b) View showing a layer. (c) Partial density of states (using a 20×2×20 half-shifted *k*-grid), showing that the VBM is primarily due to Cl *p* orbitals and CBM is primarily due to Pb *p* orbitals, as in 1D $C_8H_{28}N_5Pb_3Cl_{11}$ (Figure 4b). (d) Bandstructure, showing a smaller gap compared to the 1D structure, and an indirect gap with neither VBM nor CBM at $\Gamma$. Bands are fairly flat in the out-of-plane *y*-direction but generally dispersive in the in-plane *x*- and *z*-directions.

**TABLE S1.**

Single crystal X-ray diffraction data for $C_8H_{28}N_5Pb_3Cl_{11}$.

| Compound | $C_8H_{28}N_5Pb_3Cl_{11}$ |
|---|---|
| Formula | $C_8H_{28}Cl_{11}N_5OPb_3$ |
| Formula weight | 1221.87 g/mol |
| Temperature | 150.00(10) K |
| Crystal system | Orthorhombic |
| Space group | Pbca |
| *a* | 11.34837(8) Å |
| *b* | 15.84929(13) Å |
| *c* | 32.1489(2) Å |
| $\alpha$ | 90° |
| $\beta$ | 90° |
| $\gamma$ | 90° |
| *V* | 5782.42(8) Å$^3$ |
| *Z* | 8 |
| $\rho_{calc.}$ | 2.807 g/cm$^3$ |
| $\mu$ | 42.881 mm$^{-1}$ |
| F(000) | 4416.0 |
| Crystal size | 0.312 × 0.285 × 0.025 mm$^3$ |
| $2\theta$ range | 9.54° to 156.08° |
| Reflections collected | 56333 |
| Independent reflections | 6098 [$R_{int}$=0.1299, $R_{sigma}$=0.0435] |
| Data/restraints/parameters | 6098/1/252 |
| Goodness-of-fit on $F^2$ | 1.082 |
| Final R indexes [I>=2σ (I)] | $R_1$=0.0586 [a], $wR_2$=0.1934 [b] |
| Final R indexes [all data] | $R_1$=0.0620 [a], $wR_2$=0.1993 [b] |

[a] $R_1 = \sum ||F_o| - |F_c|| / \sum |F_o|$. [b] $wR_2 = [\sum w(F_o^2 - F_c^2)^2 / \sum w(F_o^2)^2]^{1/2}$.